\documentclass[prl,amsmath,amssymb,twocolumn,superscriptaddress,showpacs]{revtex4-1}
\usepackage{bm}
\usepackage{amssymb}
\usepackage{colordvi}
\usepackage{graphicx}
\usepackage{color}
\usepackage{hyperref}
\usepackage{ulem}
\usepackage{times}

\newcommand{\beq}{\begin{equation}}
\newcommand{\eeq}{\end{equation}}
\newcommand{\bea}{\begin{eqnarray}}
\newcommand{\eea}{\end{eqnarray}}

\begin{document}

\title{On-chip higher-order topological micromechanical metamaterials}

\author{Ying Wu}
\thanks{These authors contributed equally to this work.}
\affiliation{School of Physics and Optoelectronics, South China University of Technology, 510640 Guangzhou, Guangdong, China}

\author{Mou Yan}
\thanks{These authors contributed equally to this work.}
\affiliation{School of Physics and Optoelectronics, South China University of Technology, 510640 Guangzhou, Guangdong, China}

\author{Hai-Xiao Wang}
\affiliation{School of Physical Science and Technology, \& Collaborative Innovation Center of Suzhou Nano Science and Technology, Soochow University, 1 Shizi Street, Suzhou 215006, China}
\affiliation{School of Physical Science and Technology, Guangxi Normal University, Guilin 541004, China}

\author{Feng Li}
\email{jilinhubei@gmail.com}
\affiliation{School of Physics and Optoelectronics, South China University of Technology, 510640 Guangzhou, Guangdong, China}
\author{Jian-Hua Jiang}
\email{jianhuajiang@suda.edu.cn}
\affiliation{School of Physical Science and Technology, \& Collaborative Innovation Center of Suzhou Nano Science and Technology, Soochow University, 1 Shizi Street, Suzhou 215006, China}

\date{\today}

\begin{abstract}
Higher-order topological insulators exhibit multidimensional topological physics and unique application values due to their ability of integrating stable boundary states at multiple dimensions in a single chip. However, for signal-processing applications in high-frequency mechanical systems, the current realizations of higher-order topological mechanical materials are still limited to large-scale systems for kilohertz or lower frequencies. Here, we report the experimental observation of a on-chip micromechanical metamaterial as higher-order topological insulator for high-frequency mechanical waves. The higher-order topological phononic band gap is induced by the band inversion at the Brillouin zone corner which is achieved by configuring the orientations of the elliptic pillars etched on the silicon chip. With consistent experiments and theory, we demonstrate the coexistence of topological edge and corner states in the megahertz frequency regime as induced by the higher-order band topology. The experimental realization of on-chip micromechanical metamaterials with higher-order topology opens a regime for applications based on megahertz mechanical waves in an integrated platform where the edge and corner states act as stable waveguides and resonators, respectively.
\end{abstract}

\maketitle

{\it Introduction.}---The discovery of higher-order topological insulators (HOTIs)~\cite{Hughes2017Sci,Hughes2017prb,Hughes2019,Langbehn2017,Song2017,Schindler2018,Huber2018,Bahl2018,Imhof2018,Noh2018,Ezawa2018,AQTI,Hafezi2019,Bernevig2020qudrupole1,
Zhenbo2019quadrupole2,Xuyong2019qudrupole3,JJH2020quadrupole,JJH2020polariton,Meng2020quadrupole,LuMH2018dipole,Zhangbl2018Kagome,Khanikaev2018kagome,JJH2019Natphys,
Hassan2019,Zhangshuang2019,Christensen2019,Xiabz2019elastic,Iwamoto2019dipole,DongJW2019dipole,LuMH2019dipole,JJH2020surfacewave} opens a frontier in the study of topological phenomena and materials beyond the conventional paradigm of topological insulators and their classical wave analogs~\cite{Hasan2010,Qi2011,Ozawa2019,Zhangxj2018,Ma2019,Liu2019}. Acoustic~\cite{Zhangbl2018Kagome,Khanikaev2018kagome,JJH2019Natphys,Meng2020quadrupole} and elastic~\cite{Huber2018,Xiabz2019elastic} metamaterials, due to their excellent controllability of Bloch bands and band topology, have played important roles in the experimental realizations of HOTIs. Such classical analogs of HOTIs have demonstrated versatile mechanisms for the manipulation of waves as enabled by the coexisting multidimensional topological boundary states. Using mechanical HOTIs, it is possible to integrate stable physical channels of different dimensions (e.g., edge and corner states) in a single chip. However, so far, mechanical HOTIs remain in the kilohertz regime with large-scale building blocks~\cite{Zhangbl2018Kagome,Khanikaev2018kagome, JJH2019Natphys, Meng2020quadrupole}. To reach the regime for sensing and information-processing applications, mechanical HOTIs must be scaled down to the chip-scale level for megahertz (MHz) mechanical wave manipulations.

\begin{figure}[!h]
	\includegraphics[width=3.4in]{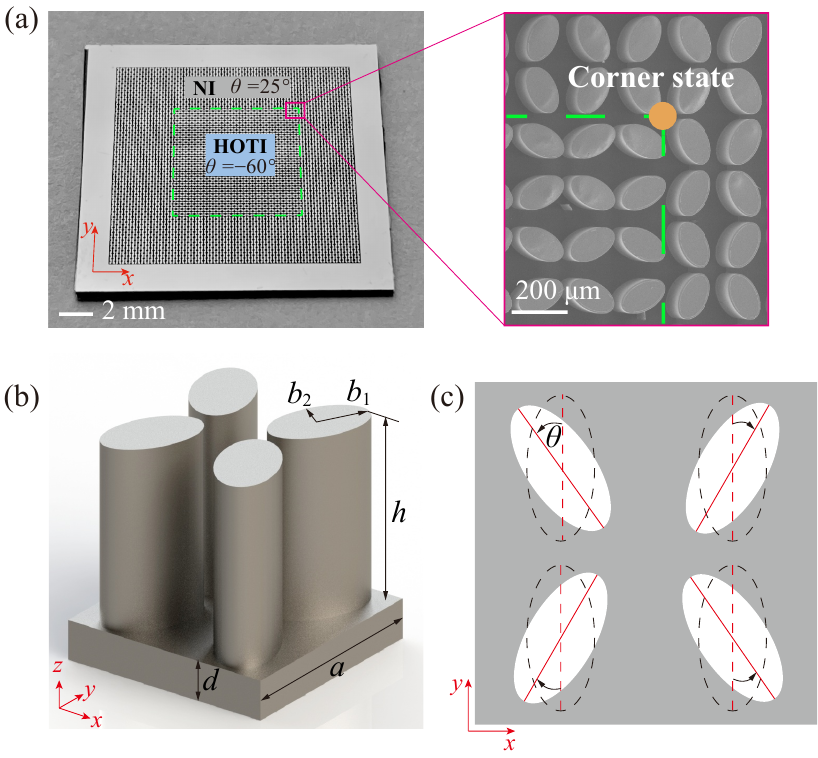}
	\caption{Mechanical HOTI on a silicon chip. (a) Photograph of the silicon chip which is etched to form square-lattice micromechanical metamaterials with 40$\times$40 unit cells. In the etched area, the region inside the green dashed box is the mechanical HOTI with $\theta=-60^\circ$ [see (c) for the definition of the orientation angle $\theta$], whereas the region outside the box is the mechanical NI with $\theta=25^\circ$. The inset shows the zoom-in of the corner region enclosed by the red box. A topological corner state emerges around the corner marked by the yellow dot. (b)-(c): The geometry and the top-down view of a unit cell, respectively. }
\end{figure}

In this Letter, we report the experimental realization of higher-order topological micromechanical metamaterials based on etch-patterning of thin silicon films that form two-dimensional square-lattice structures consisting of elliptic pillars on the substrate. The band topology is controlled by the geometry configuration of the elliptic pillars in each unit-cell. By inducing a band inversion at the Brillouin zone corner, a higher-order topological band gap is formed on the phonon branches dominated by out-of-plane polarization. The topological edge and corner states are predicted to emerge as due to the higher-order band topology. We experimentally demonstrate the presence of the edge and corner states at MHz frequencies, and characterize their dispersions through piezoelectric excitations and laser vibrometer scanning. The consistency between the theory, simulation and experiments demonstrate the multidimensional topological physics and reveal the effectiveness of our approach toward higher-order topology in micromechanical systems. The coexisting edge and corner states could be used in integrated microelectromechamical systems as stable and compact waveguides and resonators, respectively.

\begin{figure}[!h]
	\includegraphics[width=3.4in]{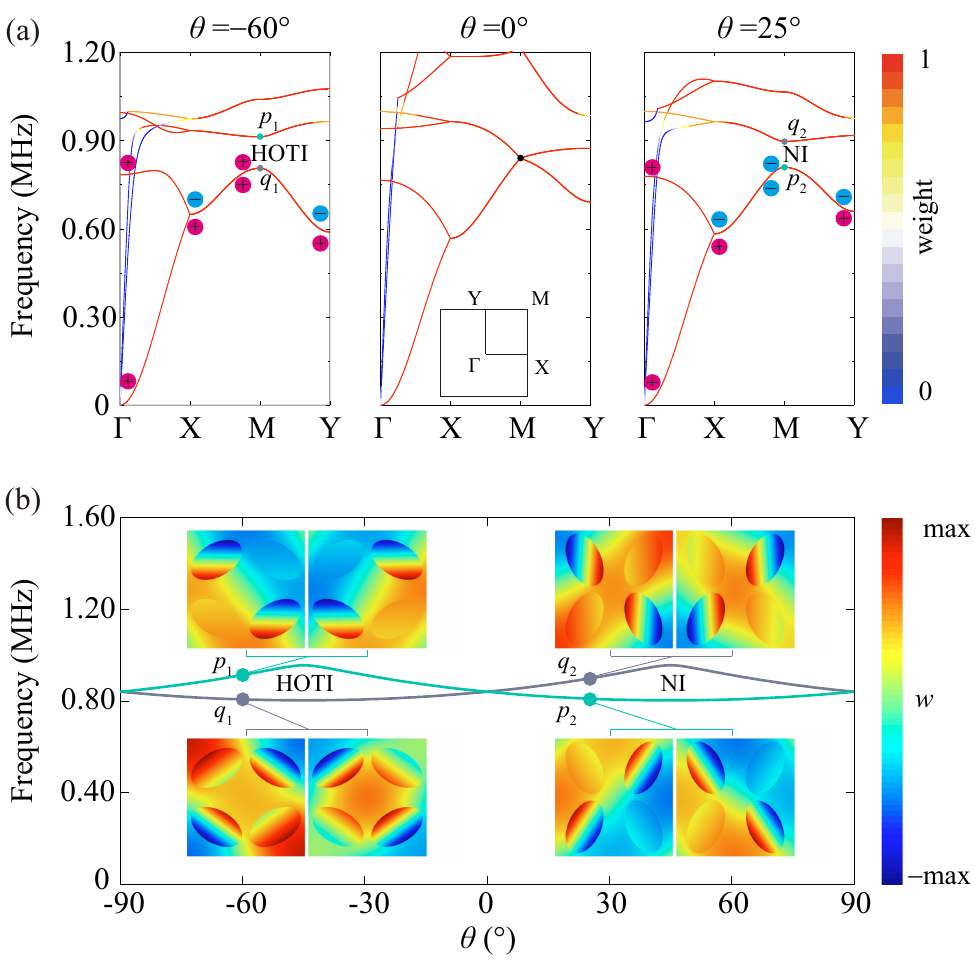}
	\caption{Phononic bands and topological band gap. (a) Phononic band structures and parity eigenvalues [$+$ (red dots) and $-$ (blue dots) denote the even and odd parities, respectively] for micromechanical metamaterials with different orientation angles $\theta=-60^\circ$ (HOTI), $0^\circ$ (gapless), and $25^\circ$ (NI). The gray and green dots denote the even- and odd-parity eigenmodes at the $M$ point, respectively. (b) The band edge frequencies at the $M$ point as functions of $\theta$. Insets show the profiles of the out-of-plane displacement $w$ in a unit cell for $\theta=-60^\circ$ and $25^\circ$, separately. }
\end{figure}

{\it On-chip mechanical HOTI.}---A sample with a mechanical HOTI surrounded by a mechanical normal insulator (NI) is fabricated by etching a silicon chip on the silicon dioxide substrate using micro-manufacturing technology (see Supplemental Material~\cite{SM}). As inspired by Ref.~\cite{JJH2019Natphys}, the mechanical HOTI and NI are realized using a unified scheme of square-lattice micromechanical metamaterials with four eliptic pillars in a unit-cell (Fig. 1). Depending on the orientation angle $\theta$ of the eliptic pillars, such a micromechanical metamaterial can realize both the HOTI and NI. As shown in Fig.~1(a), a large structure with the HOTI enclosed by the NI has edge and corner boundaries where the topological edge and corner states appear, separately. Fig.~1(b) gives the structure of the micromechanical metamaterial which has a lattice constant $a=380$~$\mu$m. The etching depth is $h=310$~$\mu$m. The thickness of the bottom silicon layer is $d=70$~$\mu$m. The semimajor (semiminor) axis of the eliptic pillars is $b_1=90$~$\mu$m ($b_2=53.2$~$\mu$m). As illustrated in Fig.~1(c), the orientation of the eliptic pillars are characterized by the angle $\theta$. The rotation centers that define the angle $\theta$ are at $(\pm\frac{a}{2}, \pm\frac{a}{2})$. Here, the origin of the coordinate system is at the unit-cell center and the $x$ axis is along the (100) crystalline direction of silicon. The orientations of the four pillars are arranged such that the system has the following mirror symmetries, $M_x:=(x,y)\to (-x, y)$ and $M_y:=(x,y)\to (x, -y)$, as well as the twofold rotation symmetry, $C_2:=(x,y)\to (-x,-y)$.

\begin{figure*}
	\includegraphics[width=6.6in]{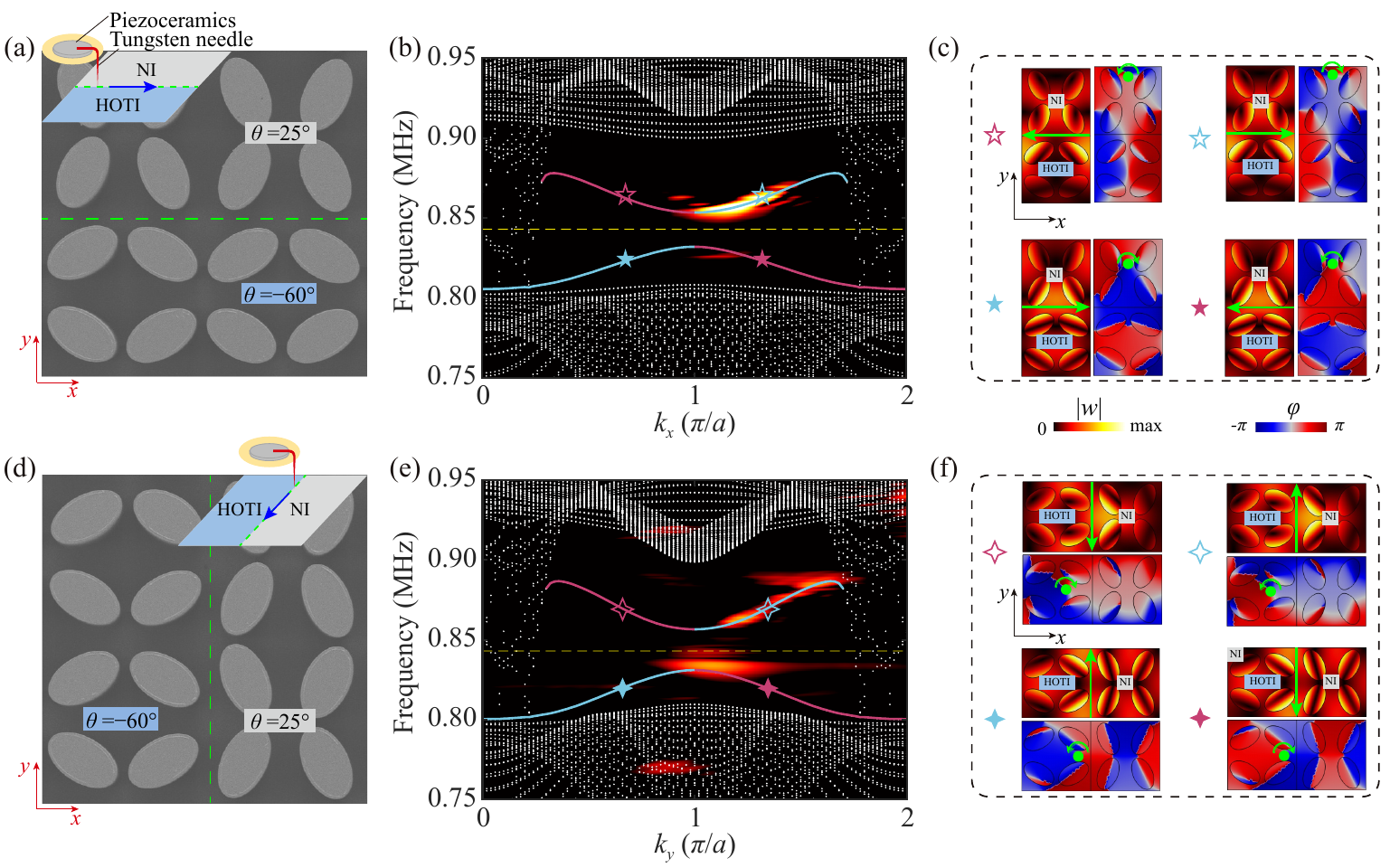}
	\caption{ Gapped edge states. (a) and (d): The edge boundaries along the $x$ and $y$ directions, respectively. The green dashed lines represent the interfaces. The insets depict schematically the experimental setups for the excitation of the edge states using a point source made of a tungsten needle and piezoceramics. The needle is placed at one end of the edge and hence excite only forward-propagting edge waves (these with positive group velocity). (b) and (e): The calculated dispersions of the edge states (red and cyan curves) and the projected bulk bands (points) as well as the measured dispersion of the forward-propagting edge states (hot color). The dashed yellow lines denote the frequency of the corner states. (c) and (f) The displacement $\left| w \right|$ and phase $\varphi$ profiles of the edge states marked by the stars in (b) and (e). The vortices in the phase profiles are marked by the arrows (phase winding directions) and the green dots (vortex centers).}
\end{figure*}

The elastic waves on the silicon chip, known as Lamb waves which consist of an out-of-plane and two in-plane polarizations (see Supplemental Material~\cite{SM}), are modulated by the periodic microstructure, leading to band dispersion of phonons. The phononic band structure is calculated numerically through full-wave simulations using COMSOL Multiphysics (see Fig.~2). In general, each eigenstate consists of displacements along the $x$, $y$, and $z$ directions, denoted by $u$, $v$, and $w$, respectively. To quantify the fraction of the $z$ direction displacement $w$, we calculate the weight $\int_{u.c.} |w|^2d{\bf r}/\int_{u.c.} (|u|^2+|v|^2+|w|^2)d{\bf r}$ where the integral is performed within a unit-cell (u.c.). The colorbar in Fig.~2(a) quantifies such a weight. It is seen from the figure that the phononic bands dominated by the $w$ displacement form a band gap between the second and the third bands. There are other phononic bands crossing the band gap but are dominated by in-plane displacements. We find that the band structure is effectively controlled by the orientation angle $\theta$. As shown in Fig.~2, the phononic band gap can be tuned by the angle $\theta$ to induce a band inversion at the Brillouin zone corner, i.e., the $M$ point. According to symmetry analysis of the phononic bands, the inverted band gap for $-90^\circ<\theta<0^\circ$ corresponds to a phononic HOTI, whereas the normal band gap for $0^\circ<\theta<90^\circ$ gives a phononic NI. The band gap closing at $\theta=0^\circ$ signifies a topological transition between the phononic HOTI and NI. We remark that although the unit-cell in Fig. 1(b) is not the primitive cell, it is the minimal unit-cell for the study of the edge and corner states presented in this work.

\begin{figure*}[!t]
	\includegraphics[width=6.5in]{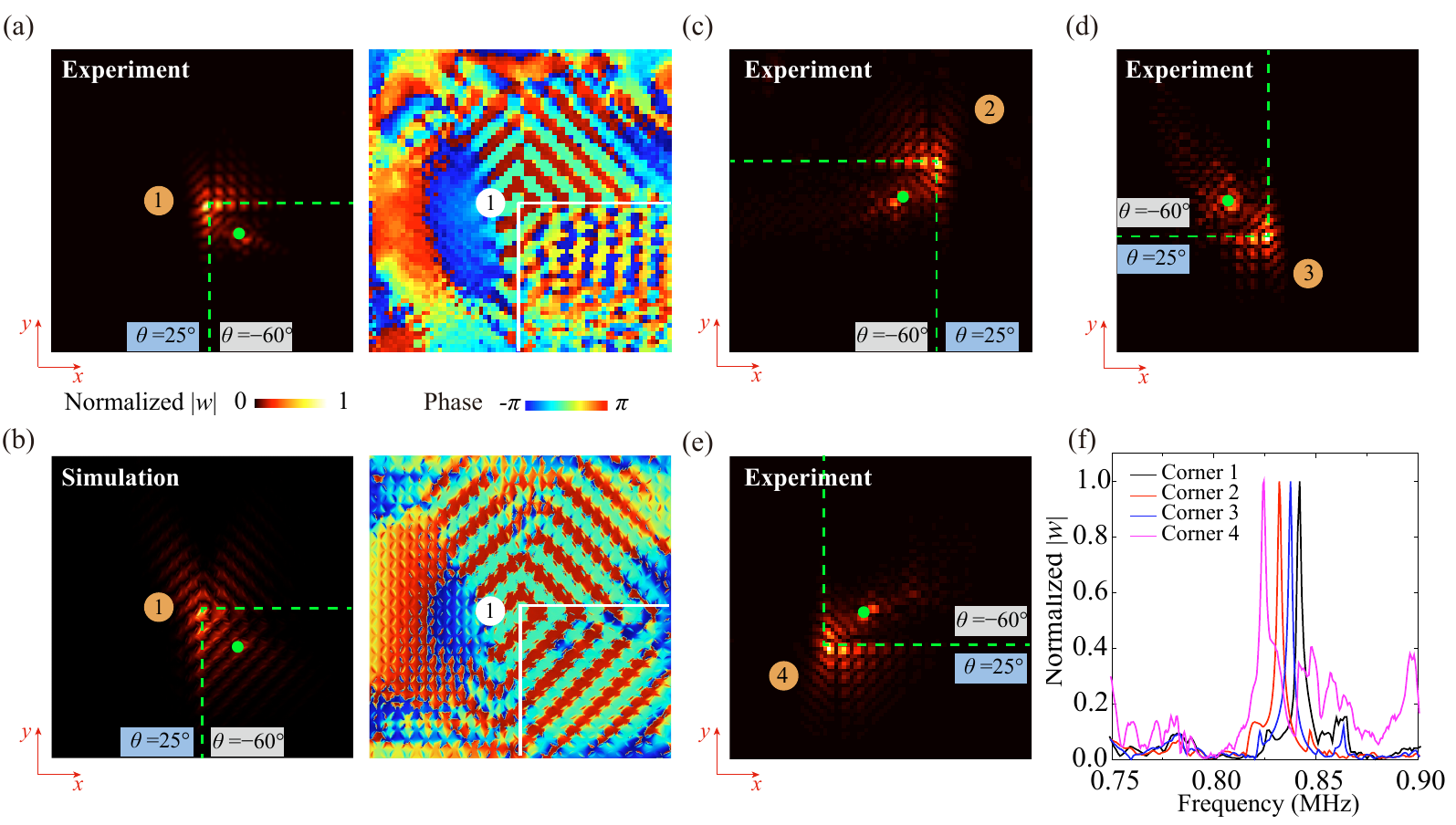}
	\caption{Measurements of the topological corner states. (a) Resonant probe of the corner state 1. Left panel: displacement profile $\left| w \right|$. Right panel: phase profile $\varphi$. These are measured at the resonance frequency of the corner 1. A point source made of a tungsten needle and piezoceramics (marked by the green dot) is placed in the HOTI region $3\sqrt{2}a$ away from the corner. (b) Simulated results of the same setup as in (a). (c)-(e) Experimentally measured displacement profiles $\left| w \right|$ of the corner 2, 3, and 4, respectively. (f) Measured source-to-detector spectra for the four corners. The detector is placed at the corner for each configuration.} 
\end{figure*}

{\it Topological index from symmetry analysis}.---The topological index of the micromechanical metamaterial can be analyzed through the symmetry eigenvalues of the two phononic bands dominated by the $w$ displacement below the band gap~\cite{Hughes2019}. The micromechanical metamaterial has $C_2$ symmetry and time-reversal symmetry which protect its band topology. Such crystalline topology is characterized by the following index~\cite{Hughes2019}, 
\begin{align}
\chi = ([X_1], [Y_1], [M_1])
\end{align}
where $[X_1]=\#X_1-\#\Gamma_1$, $[Y_1]=\#Y_1-\#\Gamma_1$, and $[M_1]=\#M_1-\#\Gamma_1$. Here, $\#X_1$, $\#\Gamma_1$, $\#Y_1$, and $\#M_1$ denote the number of even parity eigenvalues at the $X$, $\Gamma$, $Y$, and $M$ points for the two Lamb-wave phononic bands below the band gap. The higher-order band topology is characterized by the following corner charge,
\begin{align}
Q_c=-\frac{1}{4}([X_1]+[Y_1]-[M_1]) \mod 1 .
\end{align}
A fractional corner charge $Q_c$ indicates nontrivial higher-order topology~\cite{Hughes2019}. From the parities at the high-symmetry points $\Gamma$, $X$, $Y$, and $M$ indicated in Fig.~2(a) (see Supplemental Material~\cite{SM}), we find that for the HOTI, $\chi=(-1,-1, 0)$ and $Q_c=\frac{1}{2}$, whereas for the NI, $\chi=(-1,-1,-2)$ and $Q_c=0$. By constructing various boundaries between the HOTI and the NI, we can create and observe the topological edge and corner states due to the higher-order band topology.

{\it Experimental and theoretical characterizations of the edge states.}---We first study the edge states along the $x$ and $y$ directions. For this purpose, two samples are fabricated. One has edge boundary along the $x$ direction between the HOTI ($\theta=-60^\circ$) and the NI ($\theta=25^\circ$) [Fig.~3(a)], the other one has edge boundary along the $y$ direction between the two metamaterials [Fig.~3(d)]. We numerically calculate the phononic dispersion for the two samples and found gapped edge states for both of them [Figs.~3(b) and (e)]. The gapped edge states are important for stabilizing the corner states emerging in the edge spectral gap~\cite{Hughes2017Sci,Huber2018,Bahl2018}. The calculation also indicates that the edge states carry finite orbital angular momentum (OAM). As illustrated in Figs.~3(c) and (f), the phase $\varphi$ and energy flux distributions of the edge states demonstrate that the elastic energy winds around some points near the edge boundary, indicating finite OAM for the edge states. Time-reversal symmetry demands that the edge states with opposite wavevectors have opposite OAM~\cite{Lin}. Such OAM polarizations are also found in other HOTI systems~\cite{JJH2019Natphys,JJH2020quadrupole}.

In experiments, the edge states are excited by tungsten needle mounted on piezoceramic.  We measure the amplitude and phase distributions of the displacement $w$ using laser vibrometer and network analyzer (see Supplemental Material~\cite{SM}). By Fourier transforming the scanned displacement along the edge boundary, we obtain the dispersion of the edge states. As shown in Figs.~3(b) and (e), the measured edge dispersion with positive group velocity agree well with the calculated edge dispersion. In the experiments, the excitation efficiency of the lower edge branch is much smaller than the upper edge branch due to the large difference in the wavefunctions between these two branches [Figs.~3(c) and (f)]. Due to this reason, only one forward-propagating edge branch is excited and detected in the experiments.

{\it Experimental characterization of the corner states.}---To host the bulk, edge and corner states in a single chip, a box-shaped large structure illustrated in Fig.~1(a) is constructed and measured. The calculated phononic spectrum of the box-shaped large structure is presented in the Supplemental Material~\cite{SM} which indicates four nearly degenerate corner states in the spectral gap of the edge states. We find that each corner hosts a single corner state. The finite edge spectral gap stabilizes the corner states within it. We remark that the in-plane polarized bulk modes are uncoupled with the corner states and cannot be excited by the tungsten needle or measured by the laser vibrometer.

To experimentally detect and visualize the corner states, we excite each corner state using a point source in the HOTI region with $3\sqrt{2}a$ away from the corner at various frequencies. For each corner boundary, a single localized resonance is spotted in our experiments. This indicates that each corner supports only a single corner state which is consistent with the theory of HOTIs~\cite{Hughes2017Sci,Hughes2017prb,Hughes2019,xiong}. As shown in Fig.~4(a) and (b), for the corner state at corner boundary 1, the measured amplitude and phase of the displacement $w$ agree well with the simulation results at the corner resonance. The results show that the corner state is well-localized around the corner boundary and exhibit a unique spatial phase pattern. The consistency between the experimental and simulation results confirm the emergence of the corner state due to the higher-order topology.

We further measure the corner states at other three corner boundaries. The measured spatial distributions of the vibration amplitude $|w|$, as presented in Fig.~4(c)-(e), are comparable with the measured and simulated vibration amplitude distribution of the corner state 1 in Fig.~4(a). In Fig.~4(f), we present the four source-to-detector responses obtained in our experiments when the detector is set at the corners. It is visible that for each corner measurement, the source-to-detector response has only a single resonance. These resonance frequencies, 0.842~MHz, 0.837~MHz, 0.832~MHz, and 0.825~MHz, are very close to the frequencies of the corner states from eigenmode calculations $0.843$~MHz. These measurements indicate that

{\it Conclusion and outlook.}---We have achieved the experimental realization of on-chip higher-order topological micromechanical metamaterials for MHz elastic waves. These on-chip micromechanical metamaterials extend the higher-order topological phenomena into an unexplored regime and could enable applications in microelectromechanical systems~\cite{onchip1,onchip2,onchip3} with integrated waveguides and resonators (using the edge and corner states induced by the higher-order topology) as well as quantum phononics based on on-chip localized phononic modes~\cite{onchip4,onchip5}.

{\sl Note added}: We notice a preprint appearing~\cite{Yang} in parallel with our work on higher-order elastic HOTI but still in the kilohertz regime.

\begin{acknowledgments}
This work is supported by Natural Science Foundation of Guangdong Province (NO. 2020A1515010549),
China Postdoctoral Science Foundation (NO. 2020M672615), and the Jiangsu specially-appointed professor funding, China Postdoctoral Science Foundation (NO. 2020M672615, NO. 2019M662885), and National Postdoctoral Program for Innovative Talents (NO. BX20190122).
\end{acknowledgments}

\end{document}